\documentstyle[11pt,cite]{article}
\newcommand{\be}{\begin{eqnarray}}

\newcommand{\ee}{\end{eqnarray}}

\title{
	\begin{flushright}
	{\normalsize\baselineskip=10pt
        DOE/ER/40561--261--INT96--00--126\\
        TPI-MINN-96/06\\
        NUC-MINN-96/8-T\\
	May 1996 \\}
	\end{flushright}
\bf     Lattice Computations of Small-x Parton Distributions
        in a Model of Parton Densities in Very Large Nuclei
       }
\author{Rajiv V. Gavai\thanks{On sabbatical leave from the Tata Institute of
Fundamental Research, Homi Bhabha Road, Mumbai 400 005, India.  E-mail:
gavai@theory.tifr.res.in}\\ 
       {\small\it Theoretical Physics Institute, University of Minnesota, 
        Minneapolis, MN 55455} \\
        Raju Venugopalan \\
	{\small\it Institute for Nuclear Theory,
	University of Washington,
	Seattle, WA 98195--1550 } \\          
       }

\date{}

\begin{document}
\setcounter{page}{0}
\maketitle
\thispagestyle{empty}
\begin{center}
{\bf Abstract}\\
\end{center}

\noindent
Using weak coupling methods McLerran and Venugopalan~\cite{LV1} expressed the
parton distributions in large nuclei as correlation functions of a two
dimensional Euclidean field theory. The theory has the dimensionful coupling
$g^2 \mu $, where $\mu^2\sim A^{1/3}$ is the valence quark color charge squared
per unit area. We use a lattice regularization to investigate these correlation
functions both analytically and numerically for the simplified case of $SU(2)$
gauge theory.  In weak coupling ($g^2\mu L<< 5$), where $L$ is the transverse
size of the nucleus, the numerical results agree with the {\it analytic}
lattice weak coupling results. For $g^2\mu L>> 5$, no solutions exist at
O($a^4$) where $a$ is the lattice spacing. This suggests an ill-defined
infrared behavior for the two dimensional theory. A recent proposal of McLerran
et al.~\cite{Kovner} for an {\it analytic} solution of the classical problem is
discussed briefly. 

\vfill \eject

\section{Introduction}

In Ref.~\cite{LV1} McLerran and Venugopalan proposed that weak coupling
methods can be used to compute small x parton distribution functions in large
nuclei. They wrote down a partition function for wee partons with $x<<A^{-1/3}$
in the presence of external sources which are the valence quark charges. The
only large component of the valence quark current is $J^{+}$, which is modelled
by 
\be
J^{\mu}_a = \delta^{\mu +}\rho_{a}(x^+,\vec{x}_{\perp}) \delta(x^-) \, ,
\label{extcur}
\ee
where $\rho_a$ is the density (per unit area) of valence quark color charges.
Their partition function is obtained by integrating the QCD partition function,
coupled to the above static current, over all $\rho_a$'s with a Gaussian
weight. The variance of this Gaussian distribution of valence quark charges,
$\mu^2\sim A^{1/3}$ fm$^{-2}$, the average valence quark color charge squared
per unit area, is the only dimensionful parameter in the theory. If $\mu^2 >>
\Lambda_{QCD}^2$, $\alpha_S(\mu^2)<< 1$ and weak coupling methods can be used.
This model could then be studied as a toy model to understand both the rapid
growth of structure functions at small x~\cite{Lipatov} and the eventual
saturation of these structure functions as dictated by
unitarity~\cite{Froissart,Levin}. Note that the model of Ref.~\cite{LV1} is
gauge invariant due to the Gaussian distribution for the valence quark
densities.

In Ref.~\cite{LV2} the saddle point solution of the partition function in the
presence of the Gaussian random source was obtained by solving the classical
Yang--Mills equations $D_{\mu} F^{\mu\nu}= J^{\nu}$. Here $D_\mu$ is the
covariant derivative, $F^{\mu\nu}$ the non--Abelian field strength tensor and
$J^\nu$ is the current in Eq.~\ref{extcur}. It was shown that the classical
background field that satisfies the Yang--Mills equations has a simple
structure. Consequently, the classical parton distributions can be expressed as
correlation functions of a two dimensional Euclidean field theory. This is not
too surprising since it is well known that at very high energies the
longitudinal and transverse coordinates decouple. Indeed, it has been proposed
recently that the limit of $x\rightarrow 0$ and color $N_c\rightarrow \infty$
is an exactly solvable two dimensional field theory~\cite{Korchemsky}. In
papers subsequent to Ref.~\cite{LV2}, the problem of quantum fluctuations about
the background field~\cite{LV3,LV4,LV5} and that of initial conditions in heavy
ion collisions were addressed~\cite{KLW1,KLW2}. For a brief review of these
results, we refer the reader to Ref.~\cite{Venu}. An excellent introduction to
all aspects of the low x problem is given in Ref.~\cite{Kwiecinski}.

In this paper, we will discuss only the classical solutions of the Yang--Mills
equations. As we shall see in Section 2, computing the correlation functions
requires that we solve a stochastic differential equation for each color charge
configuration. Since the equations are highly non--linear, no analytic
solutions were found. However, it was claimed in Ref.~\cite{LV2} that the
parton distributions have the Weizs\"acker--Williams behavior in the weak
coupling region $\alpha_S\mu << k_t$: $dN/dx d^2k_t \propto 1/x k_t^2$. It was
conjectured that the solution of the stochastic differential equations in the
strong coupling region of $\Lambda_{QCD} <<\alpha_S \mu << \mu$ would reveal
that the classical gluons generate a screening mass $m_{screen}\sim \alpha_S
\mu$. If there is such a screening mass, it's existence would strongly suggest
that a mechanism for the restoration of unitarity at very small $x$ exists
already at the classical level. 

We will address here the question of a screening mass in the
classical theory quantitatively by solving stochastic difference equations on a
two dimensional lattice. In Section 3, we will describe how we set up the
problem and how one may use lattice perturbation theory to identify the weak
coupling and strong coupling regimes of the theory. We define ``reduced"
correlation functions of gauge fields which are one dimensional projections of
the original two dimensional fields. If a screening mass existed in the theory,
these reduced correlation functions may be expected to have a very
characteristic exponential fall off at large distances. 

For simplicity, we will consider an SU(2) gauge theory in our numerical
work. We use the conjugate
gradient method to solve the difference equations on the lattice. Details of
the numerical procedure are also discussed in Section 3. In Section 4, we
describe lattice results for the reduced correlation functions and compare them
to the results expected from lattice perturbation theory in weak coupling. It
is observed that in the weak coupling region, the numerical results
reproduce to high accuracy the results of lattice perturbation theory. However,
as one approaches the strong coupling region on the lattice, the number of
the stochastic difference equations to which
solutions can be found decreases and eventually, in
the strong coupling region, no solutions of the lattice equations exist at the
desired O($a^4$) accuracy. In Section 5, we will interpret these results and
state our conclusions.

\section{Parton distributions as correlation functions of a 2--D field theory}
\vspace*{0.3cm}

In the model of McLerran and Venugopalan, the partition function which
describes the ground state properties of wee partons with $x<<A^{-1/3}$ and
transverse momenta $q_t<<A^{1/6}$ fm$^{-1}$, is~\cite{LV1} 
\be
        Z & = & \int~ [dA_t dA_+] [d\psi^\dagger d\psi] [d\rho]
 \nonumber \\
& & \exp\left( iS +ig\int d^4x A_+(x)\delta (x^-)
\rho (x)  - {1 \over {2\mu^2}} \int d^2x_t \rho^2 (0,x_t) \right) \, .
\label{partition}
\ee
In the above, $\rho$ is the valence quark color charge density. Also, the
parameter $\mu^2\sim A^{1/3}$ fm$^{-2}$ is the average valence quark color
charge squared per unit area. Since $\mu^2$ is the only scale in the partition
function above, the coupling constant will run as a function of this
scale~\cite{LV5}. If $\mu^2 >> \Lambda_{QCD}^2$, as will be true for {\it very}
large nuclei, $\alpha_S (\mu^2) << 1$ and weak coupling methods can be used.

If we integrate over the $\rho$ fields first, we obtain an effective action
for the wee partons with non--local propagators and vertices. Instead, the
procedure followed in Ref.~\cite{LV2} was to perform the $\rho$ integrals last.
In that approach, one needs to calculate the saddle point solution of the 
action for each $\rho$ configuration to determine the classical background
field. Any physical observable, such as a correlation function, is then
obtained by evaluating it for the saddle point solution and then
averaging it over all possible $\rho$ configurations.
The saddle point solution is nothing else but the solution to the
classical Yang--Mills equations 
\be
        D_\mu F^{\mu \nu} = g J^\nu \,,
\ee
in the presence of the external source $J^\nu = \delta^{\nu +}\rho(x_t)
\delta(x^-)$. It was shown in Ref.~\cite{LV2} that the solution of
these classical equations of motion is 
\be
        A^+ & = & 0 \nonumber \\
        A^- & = & 0 \nonumber \\
        A^j & = & \theta (x^-) \alpha^j(x_t)\label{eq:trans}
\label{class}
\ee
The transverse components $A^j$, where $j=1$, 2 further satisfy the 
equations, $F_{12}=0$ and $\nabla\cdot\alpha = g\rho$,
where the latter equation follows from the Gauss' law.  Since the fields
$A^j$ are thus gauge transforms of vacuum configurations with a gauge
condition determined by the $\rho$-configuration, one can write
$\alpha_j (x_t) = {i \over g} U(x_t) \nabla_j U^\dagger (x_t)$, where $U$ is
a unitary $SU(3)$ matrix for QCD and an $SU(N)$ matrix for a theory with
$N$ colors.  Substituting for $\alpha_j$ in the gauge condition
results in the stochastic differential equation
\be         
{\vec \nabla} \cdot U {\vec \nabla} U^\dagger = -ig^2 \rho(x_t) \, . 
\label{ugauge} 
\ee 

Using the solutions of the equation above, which are the saddle point
configuration of the partition function in Eq.~\ref{partition}, one can show
that the {\it classical} correlation functions may be expressed (in matrix
representation) as correlation functions of a two dimensional Euclidean field
theory: 
\begin{eqnarray}
\langle\alpha_i^{\alpha\beta} (x_t) \alpha_j^{\alpha^\prime \beta^\prime}
(0)\rangle_{\rho} &=& {-1\over g^2}\int [d\rho]
~\bigg(U(x_t) \nabla U^\dagger(x_t)\bigg)_{\rho}^{\alpha\beta}
\bigg(U(0)\nabla U^\dagger (0)\bigg)_{\rho}^{\alpha^\prime \beta^\prime}
\nonumber \\
&\times& \exp\left( -{1 \over {2\mu^2}}\int d^2x_t \rho^a (x_t) \rho^a
(x_t) \right)/I \, ,
 \label{fluct}
\end{eqnarray}
where the Gaussian random measure 
\be
I = \int [d\rho]~\exp\left( -{1 \over {2\mu^2}}\int d^2x_t \rho^a (x_t)
\rho^a (x_t) \right) \, ,
\label{random}
\ee
is all that is left from the original partition function.
Note that the charges are highly localized in the transverse plane:
\be
\langle \rho^a (x_t) \rangle =0\,\,;\,\,\langle \rho^a (x_t) \rho^b (y_t)
\rangle_\rho = \mu^2 \delta^{ab} \delta^{(2)}(x_t-y_t) \, .
\ee
In order to ensure that the valence quark color charge
is confined to the transverse radius of the nucleus, we require that
\be
\int d^2 x_t \rho^a (x_t) = 0 \, .
\ee
This constraint was not stated in Ref.~\cite{LV2}. In the momentum space this 
condition decrees that $\rho^a(k_t=0)=0$.  Thus the $k_t=0$ mode is excluded
explicitly.

To compute the correlation function in Eq.~\ref{fluct}, we need to solve 
Eq.~\ref{ugauge} to determine $U\equiv U(\rho)$ for each $\rho$ configuration.
We were unable to find an analytic solution to this highly non--linear 
equation for all values of the coupling.
In the following and subsequent sections we will discuss 
the analytic weak coupling solution and the numerical
solutions of this equation on a two dimensional lattice.

For completeness, let us recall that the relation between distribution
functions and the correlation functions above is straightforward and is
discussed explicitly in Ref.~\cite{LV3}: 
\be
{1 \over {\pi R^2}} {{dN} \over {dxd^2k_t}} = {1\over (2\pi)^3} {1\over x} \int
d^2 x_t~ e^{ik_t x_t}~ {\rm Tr}~[\langle\alpha_i^{\alpha\beta} (x_t)
\alpha_j^{\alpha^\prime \beta^\prime} (0)\rangle]\, ,
\label{weiz1}
\ee
where the trace is over both Lorentz and color indices.

It was argued in Ref.~\cite{LV2} that the distribution function has the 
general form
\be
{1 \over {\pi R^2}} {{dN} \over {dx d^2k_t}} =
{{(N_c^2-1)} \over \pi^2} {1 \over x}~
{1 \over \alpha_S}H(k_t^2/\alpha_S^2 \mu^2)\, ,
\ee
where $H(k_t^2/\alpha_S^2 \mu^2)$ is a non--trivial function obtained by
explicitly solving Eq.~\ref{ugauge}. The effective coupling constant of
this theory was believed to be $\alpha_S \mu/k_t$ and 
that in the ``weak coupling"
limit $\alpha_S\mu << k_t$, $H(k_t^2/\alpha_S^2 \mu^2)\rightarrow 
\alpha_S^2 \mu^2/k_t^2$, recovering the Weizs\"acker--Williams result scaled
by $\mu^2$. It was also conjectured that the function $H$ would have the
form $\alpha_S^2\mu^2/(k_t^2 + M^2)$, where $M=c\alpha_S\mu$ is a 
screening mass which is a constant $c$ times the 
dimensionful scale $\alpha_S\mu$.

Interestingly, the problem formulated above is analogous to the problem of the
critical behaviour of Ising--like models coupled to a random magnetic field. As
discussed by Parisi and Sourlas~\cite{ParSour}, the partition function in that
case has a structure identical to Eq.~\ref{partition} albeit they only
discussed the case of a scalar theory. It was argued in Ref.~\cite{ParSour}
that the singular behavior of the theory near the critical point was best
described by correlation functions analogous to those in Eq.~\ref{fluct}. The
remarkable result of Parisi and Sourlas was that their scalar version of
Eq.~\ref{fluct} could be written as correlation functions of a theory which is
identical to the original theory {\it without the random magnetic fields} but
in $D-2$ dimensions, where $D$ is the dimensionality of the original theory.
This dimensional reduction is a consequence of a hidden supersymmetry of the
expression analogous to Eq.~\ref{fluct}. The gauge theory analogue of this
symmetry is nothing other than the well known BRST symmetry. Unfortunately, 
Parisi--Sourlas dimensional reduction will not apply to Eq.~\ref{fluct} 
because the analog of their scalar field is the compact field $U$ as opposed
to the gauge field $\alpha$.

Therefore, in order to test our conjecture about the existence of a screening
mass in the the strong coupling domain, we address the question numerically by
formulating the problem on a lattice.  This also enables us to define the weak
coupling limit more precisely.  As we will discuss in the following section,
the above statements about weak coupling are modified somewhat by the precise
formulation of the problem on the lattice. The effective coupling of the theory
is indeed $\alpha_S\mu/k_t$ but only for discrete multiples of $k_t = 2\pi/L$.
Here $L$ is the transverse size of the nucleus.  Correspondingly, one obtains a
discrete version of the Weizs\"acker--Williams result for weak coupling
($g^2\mu L << 5$ as we will show) by using lattice perturbation theory. 
However, since the limit $L\rightarrow \infty$ is synonymous with strong
coupling, the Weizs\"acker--Williams result of Ref.~\cite{LV2} for continuous
transverse momenta will not be recovered. 

\section{The 2--D Theory on the lattice}
\vspace*{0.3cm}

As discussed in the previous section, to compute correlation functions of the
two dimensional field theory, we need to solve stochastic differential 
equations Eq.~\ref{ugauge} for an arbitrarily large coupling.  We intend
to do this numerically by introducing a spatial lattice.  The lattice
spacing $a$ serves as an ultra-violet regulator.  Indeed, without such a
regularization the functional integrals in Eq.~\ref{fluct} are not well
defined since the correlations of the $\rho$-fields are proportional to a 
$\delta$-function.  Introducing the lattice, one sees from Eq.~\ref{random} 
that each $\rho^a(x)$ is $\mu / a$ times a Gaussian random number of
unit variance.  Approximating the circular transverse side of the nucleus 
of diameter L by a square of length L, one sees it to be a $N \times N$ 
grid of lattice points with $L = Na$.  The continuum limit consists of
taking $a \rightarrow 0$ and $N \rightarrow \infty$ such that $L$ is
held constant.  The further removal of the infra-red regulator can be
achieved by taking $L \rightarrow \infty$, which was the limit in which
the weak coupling considerations of Refs.~\cite{LV1} and ~\cite{LV2} led
to their computational scheme of the low-x parton distributions.

Using the unitarity condition on the $U$-matrices, the stochastic 
differential equation in Eq.~\ref{ugauge} can be re--written
as
\be
\left(U\nabla^2 U^\dagger - \nabla^2 U \cdot U^\dagger \right) 
         = -2 i g^2 \rho \, ,
\ee
On the lattice, finite differences replace the derivatives:
\be
\nabla^2 U^\dagger = \sum_{j=1,2} {{\left(U^\dagger(x_t+a_j)+ U^\dagger(
x_t-a_j)-2~U^\dagger(x_t)\right)}\over a^2}  + O(a^2) \, .
\ee
The labels $j=1,2$ refer to the orthonormal directions on the 
lattice and $a_j$ denotes a displacement by a single site, i.e., by
distance $a$, in the $j$th direction. 
The resultant stochastic difference equation form (to
O($a^4$) accuracy)  of Eq.~\ref{ugauge} is
\be
\left[U(x_t)\sum_{j=1,2} \left(U^\dagger(x_t+a_j)+ U^\dagger(
x_t-a_j)\right)\right] - {\rm h.c.} + 2 i g^2\mu a~\rho(x_t) = 0  \, .
\label{differ}
\ee
In the equation above, h.c. deontes hermitean conjugate and
we have scaled $\rho\rightarrow \mu\rho/a$. This has the
advantage that the Gaussian random measure defined in Eq.~\ref{random} is now
independent of the lattice spacing $a$ and the dimensionful parameter $\mu$. It
is redefined to be 
\be
\int [d\rho^a]~\exp\left(-{1\over 2}\sum_{x_t} \rho^a(x_t)\rho^a(x_t)\right) 
\, .
\ee
The rescaled $\rho$ on the lattice satisfy the following equations in
analogy with their continuum version:
\be
\langle \rho^a (x_t) \rangle =0\,\,;\,\,\langle \rho^a (x_t) \rho^b (y_t)
\rangle_\rho =  \delta^{ab}\delta^{(2)}_{x_t,y_t} \, .
\label{latrho}
\ee
The zero net color charge constraint naturally becomes a sum on the
lattice and $ \rho^a(k_t=0)=0$ is true on the lattice as well.  
The only coupling this lattice theory has is the
dimensionless $g^2\mu a$ and the scale for the theory is provided by the 
nuclear transverse size $L$.  Physical quantities can therefore be 
obtained as a function of $g^2\mu a$ or equivalently $g^2\mu L$.

In computing correlation functions on the lattice, we will find it most
convenient to study correlations of one dimensional projections of the
two dimensional gauge fields. These ``reduced" gauge fields are defined as
\be
\alpha^r_j (x_2) = {{\sum_{x_1} \alpha_j (x_1,x_2)}\over{ N }}
\, .
\label{reduced}
\ee
This one dimensional projection sets
the momentum $k_1=0$. If there exists a mass gap $M$ in the theory, then
\be
\langle \alpha^r_\mu (x_2) \alpha^r_\mu (y_2) \rangle =
{\rm A} \exp\left(-M~|x_2-y_2|\right) \, ,
\ee
for sufficiently large $|x_2 - y_2|$ (to avoid influence of excited
states, if any).  An exponential
fall off of the correlations of these reduced gauge fields would therefore be
an unambiguous signature of a mass gap in the theory.

\subsection{Weak Coupling Limit}
\vspace*{0.3cm}

Since the above mentioned function $H$, and therefore
the correlation functions we wish to obtain, have earlier been obtained
in weak coupling limit of the 2--D theory, it will be instructive to first
calculate them in the weak coupling limit on the lattice. For small
enough $g^2\mu a$, Eq.~\ref{differ} clearly admits a solution
for $U(x_t)$ which is close to the identity matrix for all $x_t$ modulo
a {\it global} gauge rotation.  Writing the matrices $U$ in terms of the
the generators $\tau^k$ of the $SU(N)$ gauge group as $U(x_t) = \exp
(i g^2 \mu a \cdot \phi(x_t))$, with $\phi(x_t) = \sum_k \phi^k(x_t) \cdot
\tau^k$, one sees that weak coupling implies that $g^2\mu a \cdot \phi<<1$. 
One can therefore expand the field $U$ as 
\be
U = 1+ig^2\mu a \cdot \phi-{1\over 2}(g^2\mu a)^2 \cdot\phi^2+\cdots 
\ee
Keeping terms of only lowest order, Eq. ~\ref{differ} becomes
\be
\sum_{j=1,2} \left[\phi(x_t+a_j)+ \phi(x_t-a_j) - 2\phi(x_t) \right]
 = \rho(x_t)  \, .
\label{pert}
\ee
Thes equations can be solved by Fourier transforming the fields $\phi$
and the sources $\rho$ : Let
\be
\phi(x_t)={1 \over N^2} \sum_{\vec{n}} \exp\left(2\pi i
\vec{n}\cdot \vec{x_t}/L\right) \tilde{\phi}\left({2\pi\vec{n}\over L}\right)
\, , 
\ee
with $\vec{n} = (n_1, n_2)$ and $-(N-1)/2 \le n_j \le (N-1)/2$
(assuming $N$ to be odd).  Defining similarly the Fourier
transform of $\rho(x_t)$, we obtain the following solution of
the Eq.~\ref{pert}: 
\be
\tilde{\phi}\left({2\pi\vec{n}\over L}\right)={\tilde{\rho} ({2\pi\vec{n}\over
L}) \over {2 \sum_{j=1,2}\left[\cos({2\pi n_j a\over
L})-1\right]}} \, . 
\ee
Substituting back in the equation for $\phi(x_t)$, we have 
\be
\phi(x_t)={ 1 \over {2 N^2}}\sum_{\vec{n}}
{{\exp\left(2\pi i\vec{n}\cdot \vec{x_t}/L\right)}\over
{\sum_{j=1,2}\left[\cos({2\pi n_j a\over
L})-1\right]}}\tilde{\rho}\left({2\pi\vec{n}\over L}\right) \, . 
\ee

In this leading order of weak coupling,  
$\alpha_j(x_t) = g \mu \left[\phi(x_t + a_j) -\phi(x_t - a_j) \right]$, 
and the one dimensional
projections of the $\alpha$ fields, are easily computed to be 
\be
\alpha^r_1(x_2) &=& 0 \, . \nonumber \\
\alpha^r_2(x_2) &=& {g\mu i\over 2N^2} \sum_{n_2}{}^\prime 
{{\sin\left({2\pi n_2 a\over L} \right) 
\tilde{\rho} \left({2\pi n_2 a\over L}\right)}\over {\left[\cos\left({2\pi n_2 a
\over L}\right)-1\right]}}\exp\left({2\pi i n_2 x_2\over L}\right) \, .
\ee
Here prime denotes the exclusion of the  $n_2=0$ due to the total vanishing
charge condition.
To obtain the ``reduced" correlators, we take the product of the $\alpha^r$
fields and take the average over the $\tilde{\rho}$ fields. 
Using the relation between $\rho$ and $\tilde{\rho}$, and the Eq.~\ref{latrho},
one can easily show that $\tilde{\rho}$ satisfies similar equations as
well except that its two point correlation function has an extra factor
of $N^2$:
\be
\langle \tilde{\rho^a} (k_t) \tilde{\rho^b} (l_t)
\rangle_{\tilde{\rho}} = N^2 \delta^{ab}\delta^{(2)}_{k_t,l_t} \, .
\label{latrhok}
\ee
Using the relation above and after some simple algebra, we obtain
\be
\langle \alpha^r_1(x) \alpha^r_1(x^\prime)\rangle &=& 0 \, .\nonumber \\
\langle \alpha^r_2(x) \alpha^r_2 (x^\prime) \rangle
&=& {g^2\mu^2\over 2
N^2}\sum_{n_2=1}^{(N-1)/2}  {\sin^2 \left({2\pi n_2 a\over L}\right)\over
\left[\cos\left({2\pi n_2 a\over L}\right)-1\right]^2}~\cos\left({2\pi n_2 (x-
x^\prime)\over L}\right) \, .
\label{latpert}
\ee
In the continuum limit of $a\rightarrow 0$ and $N \rightarrow \infty$,
\be
\langle \alpha^r_2 \alpha^r_2 \rangle_{a\rightarrow 0} 
= {g^2\mu^2\over {2\pi^2}} 
\sum_{n_2=1}^{\infty} {\cos\left({2\pi n_2 (x-x^\prime)\over L}\right)\over 
n_2^2} \, .
\label{latcont}
\ee
By constructing similar ``reduced" correlators for the calculations of
Ref.~\cite{LV1}, one can easily see that our results are very similar to 
theirs, except that our expression above still has a finite L.  
Consequently, only discrete momenta are allowed in our sum and the lowest 
allowed momentum is $2 \pi /L$.  A naive $ L \rightarrow \infty$ yields
identical results to those of Ref.~\cite{LV1} but it turns out that this
limit is not allowed. 

In order to see why it is so, it is necessary to go back to
the weak coupling condition $g^2 \mu a \cdot \phi<<1$. Using the solution for
$\phi(x_t)$ obtained above, one can translate this condition into 
\be
g^2\mu L \left\{{1\over N^3}\sum_{\vec{n}} {{\exp\left(2\pi
i\vec{n}\cdot \vec{x_t}/L\right)}\over {\sum_{j=1,2}\left[\cos({2\pi n_j
a\over L})-1\right]}}\tilde{\rho}\left({2\pi\vec{n}\over L}\right)\right\}~<<~1
\, . 
\ee
Taking further the continuum limit, one obtains  
\be
g^2\mu L \left\{{1\over {4\pi^2 L}}\sum_{\vec{n}} {{\exp\left(2\pi
i\vec{n}\cdot \vec{x_t}/L\right)}\over {j_1^2 +j_2^2} }
\tilde{\rho}\left({2\pi\vec{n}\over L}\right)\right\}~<<~1
\, . 
\ee
One can now see that the $L \rightarrow \infty$ limit will violate the above
condition even if one ignores the possibly logarithmically divergent 
factor in the curly bracket in that limit.  The weak coupling condition
thus constrains $L$ to stay finite and small.
The expression in the curly brackets can be evaluated numerically and typically
the largest values are $\sim 0.2$ if one keeps $L$ finite.
The weak coupling condition for a finite size $L$
is then (approximately)
\be 
g^2\mu L << 5 \, .
\ee

The correlation function $\langle \alpha^r_1 \alpha^r_1 \rangle$ 
becomes non-zero in the next-to-leading order of the expansion, when
\be
\alpha_\mu = g\nabla_\mu \phi + ig^3 \left(\phi\nabla_\mu \phi-(\nabla_\mu 
\phi)\phi\right) \, .
\ee
For the sake of brevity we have used here the continuum notation to denote the
finite differences. Using this expression, the reduced correlator can be
computed straightforwardly. The final expression is fairly tedious (involving
the Gaussian average of four $\tilde{\rho}$ fields). The key result of this
computation is that in the continuum limit $a\rightarrow 0$, 
\be
\langle \alpha^r_1 \alpha^r_1 \rangle\, 
\propto g^2\mu^2 (g^2\mu L)^2 \, ,
\label{transverse}
\ee
i.e., the correlation function grows as $(g^2\mu L)^2$. In the next section, we
will compare the results of our lattice computation with these lattice
perturbation theory results in the weak coupling region $g^2\mu L << 5$. 

The weak coupling calculations above were done by introducing an
ultra-violet cut-off, the lattice spacing $a$. Since the final result
shows sensitivity only to the infra-red regulator, namely the size $L$, 
one can ask whether these results can be derived without introducing the 
lattice in $x$-space at all.  The answer turns out to be affirmative.
One can easily show that the continuum problem can be formulated in the
momentum space of a finite square box of length $L$.  Due to the 
discrete momentum spectra, the corresponding $\tilde{\rho}$-measure
is then well defined. One then solves Eq.~\ref{ugauge} by first
expanding $U(x)$ and then Fourier transforming the resultant
equation.  The final results, of course, remain unchanged when 
compared with the $a \rightarrow 0$ limit above.

\subsection{Numerical Method}
\vspace*{0.3cm}

In order to obtain a result for the correlation functions of the
$\alpha^r$-fields which is free of the infra-red cut-off, one has
to take the limit $L \rightarrow \infty$.  Since it thus necessarily
takes one out of the weak coupling region, 
we now turn to the procedure we used to solve the stochastic difference
equations in Eq.~\ref{differ} numerically.  To simplify our
computations and as a test, we choose to work with the
gauge group $SU(2)$.  No qualitative differences are anticipated with
regard to the existence of the mass gap as a result of this
simplification.  Writing an SU(2) matrix $U$ as $ a_0 {\rm I} + i \tau^k
a_k$, we can write the first term of Eq.~\ref{differ} as
\be
\left[U(x_t)\sum_{j=1,2} \left(U^\dagger(x_t+ a_j)+ 
U^\dagger(x_t-a_j)\right)\right] = b_0 {\rm I} + i\tau^k b_k \, .
\ee
Here $I$ is the unit 2$\times$2 matrix and $\tau^k, k=1,2,3$ are the Pauli
matrices. The coefficients  $a_k$ satisfy the unitarity
condition, $\sum_{k=0}^3 a_k^2 = 1$ but the coefficients $b_k$ do not.
Eq.~\ref{differ} can now be re-expressed as $b_k(x_t, x_t \pm a_j) +
g^2 \mu a \cdot \rho_k(x_t) = 0$.  In order to solve these coupled nonlinear
equations, we minimize the function $F$, defined by
\be
F &=& \sum_{x_t} \Bigg\{\sum_k \left(b_k(x_t, x_t \pm a_j) +g^2\mu
a~\cdot \rho_k(x_t)\right)^2 +
\left( \sum_k a_k^2(x_t)-1\right)^2\Bigg\} 
\, .
\ee
Minimizing $F$ is equivalent to solving Eq.~\ref{differ}
for each lattice point and color charge ($3~N^2$ equations) while simultaneously
imposing the unitarity condition $U~U^\dagger =1$ at each point on the lattice.
The latter is done by the second set of terms in $F$.  Note that $F$ is
a sum of squares of real numbers with zero as its possible absolute and
desired minimum.  A lack of solution will be signalled by large
values of $F_{\rm min}$ for the absolute minimum.

We use a multi-dimensional conjugate gradient method, described in the 
subroutine $FRPRMN$ and its associated subprograms in Ref.~\cite{NR},
to minimize $F$ to an accuracy better than O($a^4$) as dictated by
accuracy of the original lattice equations.  The zero net charge
condition compels us to use periodic boundary conditions in accordance
with the Gauss' law.  We investigated both ordered and random starts for
the initial guesses for the $U$'s.  Each iteration consisted
of choosing the source distributions randomly over the entire lattice in
the momentum space such that 1) $\tilde{\rho}(0,0) = 0$, 2)
$\tilde{\rho}^*(\vec{k}) = \tilde{\rho}(-\vec{k})$ and 3) both the real
and imaginary parts of each $\tilde{\rho}(\vec{k})$ were random Gaussian
numbers with variance $1/\sqrt{2}$.  The $\rho$-distribution was then
obtained by an explicit Fourier transformation.  Using the conjugate
gradient method, the set of $U$'s for the absolute minimum was 
found.  If the minimum was $O(a^4)$ or smaller, then 
the matrices $U\equiv U(\rho)$ were used to compute the correlation functions
on the lattice. We obtain $\alpha_\mu$ from the relation
\be
\tau_k\cdot\alpha_\mu^k = -{1\over ga} {\rm Im} \left( U(x_t)U^\dagger
(x_t+a_\mu) \right)\, .
\label{lalp}
\ee
We have also checked that the symmetric difference definition
for the derivative yields the same result.
Just as in Eq.~\ref{reduced}, we define the ``reduced" gauge fields 
$\alpha^r$ and compute correlators by taking the product of these
gauge fields.  This procedure is repeated over several iterations,
typically a few hundred, and the the correlation function is averaged
over these iterations.  The errors are determined in the usual way by
computing the standard deviations.  Note that the sets of $\rho$'s in
successive iterations are totally independent and one thus has negligibly 
small auto-correlations.

\section{Results}
\vspace*{0.3cm}

In view of the facts that that the coupling for the lattice theory above
is $g^2\mu a$, and that none of the three quantities in this expression
occur independently, we chose to set $g^2\mu=1$ in our simulations and varied 
$g^2\mu a$ by varying the lattice spacing $a$. 
Simulations were performed for a range of lattice sizes,
ranging from $N=21$ to $N=211$, and for values of
$g^2\mu L$ ranging from 0.5 to 20. Typically 200 iterations were performed,
each consisting of an independent set of the $\rho$-distributions,
unless stated otherwise.  Noting from Eq.~\ref{latpert} that 
$g^2 \mu^2$ sets the scale of the correlation functions, and using the 
definition in Eq.~\ref{lalp}, one can show that the factor
$(g^2\mu a)^2$ relates the dimensionless lattice correlation function to
the physical $\alpha^r$-correlations.  We therefore show the results for
the latter in the units of $g^2 \mu^2$.

In Fig.~1a we show the results of our computation for $\langle \alpha^r_j(x)
\alpha^r_j(0) \rangle /g^2\mu^2$, $j=1$, 2, plotted as a function of the
dimensionless distance $x/L$ for a small value of $g^2 \mu L =0.5$. The lattice
size was $41 \times 41$. Also shown are the analytic weak coupling results of
Eq.~\ref{latpert} for this lattice size which agree rather well with the direct
computation . One also sees clearly that the $\alpha^r_1$-correlation is very
small compared to the $\alpha^r_2$-correlation.  In fact, the former is
consistent with zero on the scale of this plot.  This, together with the
excellent agreement with the weak coupling result, reassures us that 1) for
small $g^2 \mu L$ the assumptions made in deriving the weak coupling results
are indeed justified and 2) our numerical procedure works fine.  

Fig.~1b further shows that these results are indeed the continuum results.  It
displays the results for the same $g^2 \mu L$ but on $N=21$ and 41 lattices.
The results for the $j=2$ correlation are again displayed as a function of the
dimensionless distance $x/L$ and the results are seen to be lattice size
independent.  One may wonder why the correlation function is negative, given
that it can now be thought of as a continuum property.  The weak coupling
result of Eq.~\ref{latpert} provides a hint for understanding this. The leading
term in it is negative for $x-x^\prime=L/2$ and the successive terms alternate
in sign and become progressively smaller in magnitude. Thus for any finite $L$,
the correlation function will be negative midway if one is in the weak coupling
domain. 

Although the $\langle \alpha^r_1(x) \alpha^r_1(0) \rangle$ 
correlation function
appears to be zero at all $x/L$ in Figs.~1a and 1b, it has an interesting
structure as well. As Fig.~2 shows for $N=21$ and 41 lattices, this
correlation function decreases monotonically as $x$ increases but
remains positive all through.  We will later compare this behavior for
larger $g^2 \mu L$ but it is interesting to note this difference with
the leading order weak coupling behavior.   As remarked earlier in
Section 3.1, we do expect a non-vanishing contribution to it from the
next-to-leading order contribution and a detailed examination of it also
reveals it to be positive definite.

Having tested both the weak coupling limit and the conjugate gradient method on
the lattice, we increased the $g^2 \mu L$, first by retaining the same lattice
size of $N=21$ and then increasing it as well up to $N=71$ such that the
lattice spacing stayed at $a \simeq 0.1$.  This value was determined by making
runs on the $N=21$ lattice for various $a$ and by checking that the errors due
to finite $a$ remained small.  For the rest of our numerical work we have
attempted to stay close to this value of $a$; increasing thus the lattice size
$N$ in order to increase $g^2 \mu L$.   Fig.~3 displays the results for the
$\alpha^r_2$-correlation functions in the units of $g^2 \mu^2$ as a function of
$x/L$ for $g^2\mu L=$~0.5,1, 2, 3, 4, 5, 6, and 7.   It appears to remain
almost independent of $g^2 \mu L$ until it reaches our estimated region of the
validity of the weak couping theory: for $g^2 \mu L \ge 5$ the correlation
function tends to be less and less negative as $g^2 \mu L$ increases.  This
signals a departure from the leading order weak coupling result which, as seen
in Eq.~\ref{latcont}, is independent of $g^2 \mu L$ when viewed as a function
of the dimensionless variable $x/L$.  An obvious source of the departure from
Eq.~\ref{latpert} are higher order contributions.  If these tedious terms are
indeed responsible for it then one expects a growth in the
$\alpha^r_1$-correlation as we argued in Section 3.1.  

Fig.~4 exhibits the $\alpha^r_1$-correlation function in the units of $g^6
\mu^4 L^2$ as a function of $x/L$ for $g^2 \mu L$ up to 5.  They do indeed
group together to suggest a universal curve, and thus confirm the rise of this
correlation function as $(g^2 \mu L)^2$.  Fig.~5 demonstrates this in another
way and also suggests that $g^2 \mu L \sim 5$ is the boundary of the weak
coupling region.  What is shown there is the $x=0$ value for this correlation
function in the units of $g^2 \mu^2$ as a function of $g^2 \mu L$, both before
and after scaling out the factor $(g^2 \mu L)^2$. Note the scale of both the
axes. A linear rising curve is thus an indication of the power law which seems
to be consistent with the power two.  What can also be inferred from this
figure is a small trend to push this power up as one goes above $g^2 \mu L
\simeq 5$.  A priori, such a behavior could also be due to yet more higher
order terms. However, these results also suffer from a further defect. 

For the larger values of $g^2 \mu L$, one sees the $F_{\rm min}$ slowly creep
up and go beyond the O$(a^4)$ level.  Indeed, typically one fails to obtain any
acceptable minimum at that level for about 10-15 \% of the iterations.   This
should be contrasted with the small $g^2 \mu L$ case where $F_{\rm min}$ was a
lot smaller than O$(a^4)$ for {\it each} iteration.  Increasing the $g^2 \mu L$
even further, this becomes worse very quickly and by $g^2 \mu L = 10$ no
minimum exists at that accuracy.  

In order to better understand the reason behind this, we show in Fig.~6 a
normalized histogram plot for $g^2 \mu L=6$, 7, 10 and 20. These runs were made
on $N=61$, 71, 111 and 211 lattices and the latter two have very few
iterations, being 11 and 6 respectively.  All the corresponding minima were too
high compared to O($a^4$).  The parameter $R$ in Fig.~6 is defined as follows.
Defining $R^k(x_t)$, $k$=1, 2 and 3, to be the terms on the LHS of
Eq.~\ref{differ} divided by $a^4$, one sees that $R^k(x_t) =0$, for all $x_t$
and $k$, is the desired solution. A value of $R^k(x_t)\neq 0$ measures how far
away from the desired minimum (found by minimizing $F$) is the solution for
that value of $x_t$ and $k$. What Fig.~6 depicts, for different values of $g^2
\mu L$, is the fraction of the $3N^2$ equations which have, for the best
minimum of $F$, $R$ given by the value on the $x$-axis. We have checked that
the similar histogram plots for the weak coupling region have only the bin near
zero occupied, i.e. they peak sharply at zero. What one sees in Fig.~6 though,
are increasing deviations away from $R=0$ -- fewer and fewer of the 3$N^2$
equations are being satisfied at the required level of accuracy. Noting that
$R=1/a$, which is $\sim 10$ for these runs, corresponds to the equations not
being satisfied at O($a^3)$ level, one finds that the minimum of $F$ has
increasingly many equations like that. This is thus an indication that for
$g^2\mu L \ge 10$, no solutions to the stochastic equations exist at O($a^4$). 

Our results therefore suggest the following: when $g^2\mu L < 5$,  the 
correlation functions computed directly agree very well with the
expectations from lattice perturbation theory. For intermediate values,
$g^2\mu L\approx 5$, the lattice results still agree reasonably well with
the analytical lattice expressions but one notices an increasing
trend of lack of solutions to more and more equations
at the $a^4$ level. For larger values of $g^2\mu L > 10$, no
solutions exist at that level. 

The absence of solutions as we increase $g^2\mu L$ was unexpected. Increasing
$g^2\mu L$ for a fixed value of $a$ is equivalent to merely increasing the
number of sites $N$. In other words, the number of equations has been increased
but the structure of the equations and the coupling is unchanged. Why then are
there no solutions as we go beyond $g^2\mu L\sim 5$? One way to understand this
is as follows: since $U=\exp(ig^2 \mu a \cdot \phi)$ and in weak coupling 
$\phi~\propto~ N$, increasing $g^2\mu L$ will cause the $U$ matrices to deviate
increasingly away from identity. However, since $g^2\mu a$ is unchanged and
it remains small, and since $\rho$ remains O(1), Eq.~\ref{differ} will still 
prefer the $U$ to be close to identity.
The ensuing mismatch will thus result in a lack of solutions
for large $g^2\mu L$. It is possible that our lack of solutions is because
our boundary conditions are too restrictive. However, because we need to satisfy
Gauss' law in two dimensions, periodic boundary conditions appear to be the 
appropriate physical choice.

The absence of solutions for $g^2\mu L~>~5$ is a serious problem for the
classical theory discussed in Ref.~\cite{LV2}. Not only because the
conjectured scenario of a screening mass needed the coupling to be
strong but also because the removal of the infra-red regulator pushes
one in that region.  Considering that even the very large nuclei will be
finite in size, one could check whether the condition above is
physically acceptable.  If we take $L\approx 2~A^{1/3}$,
then the weak coupling condition $g^2\mu L~<~5$, holds only for very small $A$.
This is due to the fact that $\mu \sim A^{1/6}$ and the coupling $g^2$
is also evaluated at the scale $\mu$.  Thus
this condition contradicts the weak coupling assumption of the four dimensional
theory in Ref.~\cite{LV1} which is expected to be valid only for very large
$A$. 

\section{Summary and Outlook}
\vspace*{0.3cm}

In Ref.~\cite{LV1}, a QCD based model was formulated to study the properties of
low x, wee partons in large nuclei. For very large nuclei, it was argued that
the problem could be formulated as a weak coupling, many body problem. In
Ref.~\cite{LV2}, it was shown that the classical saddle point solution of the
model could be expressed as a two dimensional Euclidean field theory with the
dimensionful coupling $\alpha_S\mu$. Computing correlation functions in this
2--D theory required the solution of highly non--linear stochastic differential
equations in the presence of a Gaussian random source. No analytic solution of
these equations was found in Ref.~\cite{LV2}. 

However, it was argued that classical distribution functions had a
Weizs\"acker--Williams distribution at large momenta $k_t >> \alpha_S\mu$. It
was also conjectured that at smaller momenta, in ``strong coupling", the theory
acquired a screening mass $M\sim \alpha_S\mu$, which regulated the growth of
the distribution function at small $k_t$. The existence of a screening mass
would be suggestive of a weak coupling, albeit non--perturbative, restoration
of unitarity already at the classical level.

In this paper, we have discussed the analytic weak coupling and the
numerical solutions of the stochastic differential equations on a two 
dimensional lattice. For our numerical work, we made the simplifying
assumption of two colors and investigated an $SU(2)$ gauge theory.
With lattice perturbation theory as our
guide, we identify $g^2\mu L << 5$ as the weak coupling condition. Our
numerical results on the lattice agree very well with lattice perturbation
theory for these values. For larger values of $g^2\mu L$, no solutions are
found which satisfy the stochastic equations at the required level of accuracy.
Thus not only is a screening mass absent but a further implication of this
result is that the classical theory is ill defined in the infrared.
Furthermore, if we identify $L\sim 2~A^{1/3}$ fm, the lattice weak coupling
condition is satisfied only for very small A. This limit appears to contradict
the weak coupling limit in the full theory, which is expected to be valid only
for very large nuclei. In sum, our work suggests that the classical
theory in Ref.~\cite{LV2} is seriously flawed. 

Recently, McLerran and collaborators~\cite{Kovner} have proposed that the
original classical theory is flawed because the authors in Ref.~\cite{LV2}
failed to properly solve the Yang--Mills equations for the transverse
components $A^i$ of the classical field. These are determined through the
equation 
\be
\nabla_i \partial^+ A^i + A_i\times \partial^+ A^i = g J^+ \, .
\ee
Ref.~\cite{LV2} argued for a solution of the form in
Eq.~\ref{class}: $A^i=\alpha^i \theta(x^-)$. If one then ignores 
the commutator terms, because it involves fields at the 
same $x^-$, one then obtains $\nabla\cdot
\alpha=g\rho$, which gives us the stochastic equation we solved for $U$ using
$\alpha_i =U\nabla_i U^\dagger/(-ig)$. The authors of Ref.~\cite{Kovner} argue
that the cross product term above cannot be dropped because of its peculiar
singular structure. They argue that the source term must be regularized so that
instead of being a $\delta$-function in $x^-$, the charge density 
$\rho$ depends on the spacetime rapidity $y=-\log(x^-)$. The above equation
is then re--written as
\be
D_i {dA^i\over dy} = g \rho(y,x_t) \, ,
\ee
where $D_i$ is the covariant derivative. It is claimed in Ref.~\cite{Kovner}
that this equation can be solved exactly and the correlation functions computed
analytically. The distribution functions have the Weizs\"acker--Williams form
for large transverse momenta, $dN/d^2 k_t\sim 1/k_t^2$. For small transverse
momenta, it has the logarithmic form $dN/d^2 k_t \sim
\log(k_t^2/\chi(y,k_t^2))$. Here $\chi(y,k_t^2)=\int_{max_{y,y^\prime}}^ {y_0}
dy^\prime \mu^2(y^\prime,Q^2)$. We refer the reader to Ref.~\cite{Kovner} for
the details of their calculation.

\section*{Acknowledgments}

We would both like to thank Jamal Jalilian--Marian, Alex Kovner, Larry McLerran
and Heribert Weigert for useful discussions. We would also like to thank the
Theoretical Physics Institute, University of Minnesota and the Institute for
Nuclear Theory, University of Washington for hospitality to R.V and R.V.G
respectively. Research supported by the U.S. Department of Energy under grants
No. DOE Nuclear DE--FG06--90ER--40561 and  DE-FG02-87ER40328.

\newpage

\section*{Figure Captions}

\noindent Figure 1: 
(a) The correlation functions $\Gamma^j(x) \equiv \langle \alpha^r_j(x)
\alpha^r_j(0) \rangle$ as a function of $x/L$ for $j=1$ (circles) and
$j=2$ (crosses) for $g^2 \mu L = L = 0.5$ and $L = 41 a$. The continuous
line is the weak coupling result of Eq.~\ref{latpert}.  (b) Same as
Fig.~1a but for $L=21a$ and $L=41a$ and for $\Gamma^2$ only.
\bigskip

\noindent Figure 2: 
The $\alpha^r_1$-correlation function as a function of $x/L$ for $g^2
\mu L =0.5$ and $N=21$ and 41 lattices.
\bigskip

\noindent Figure 3: 
The $\alpha^r_2$-correlation function as a function of $x/L$ for $g^2
\mu L =0.5$ and 1, 2, 3, 4, 5, 6, and 7. The lattice sizes can be found
in the text.
\bigskip

\noindent Figure 4: 
The $\alpha^r_1$-correlation function in the units of $g^6 \mu^4 L^2$
as a function of $x/L$ for $g^2 \mu L =0.5$ and 1, 2, 3, 4, and 5. 
The lattice sizes can be found in the text.
\bigskip

\noindent Figure 5: 
The $\alpha^r_1$-correlation function at $x=0$ as a function of $g^2 \mu
L$ without (crosses) and with (squares) a division by $(g^2 \mu L)^2)$.
\bigskip

\noindent Figure 6: 
The histograms of the fraction of total equations solved at an accuracy 
$R*a^4$ as a function of $R$. These are plotted for the following values of
$g^2\mu L$: 6, 7, 10 and 20. For details, see the text.
\bigskip

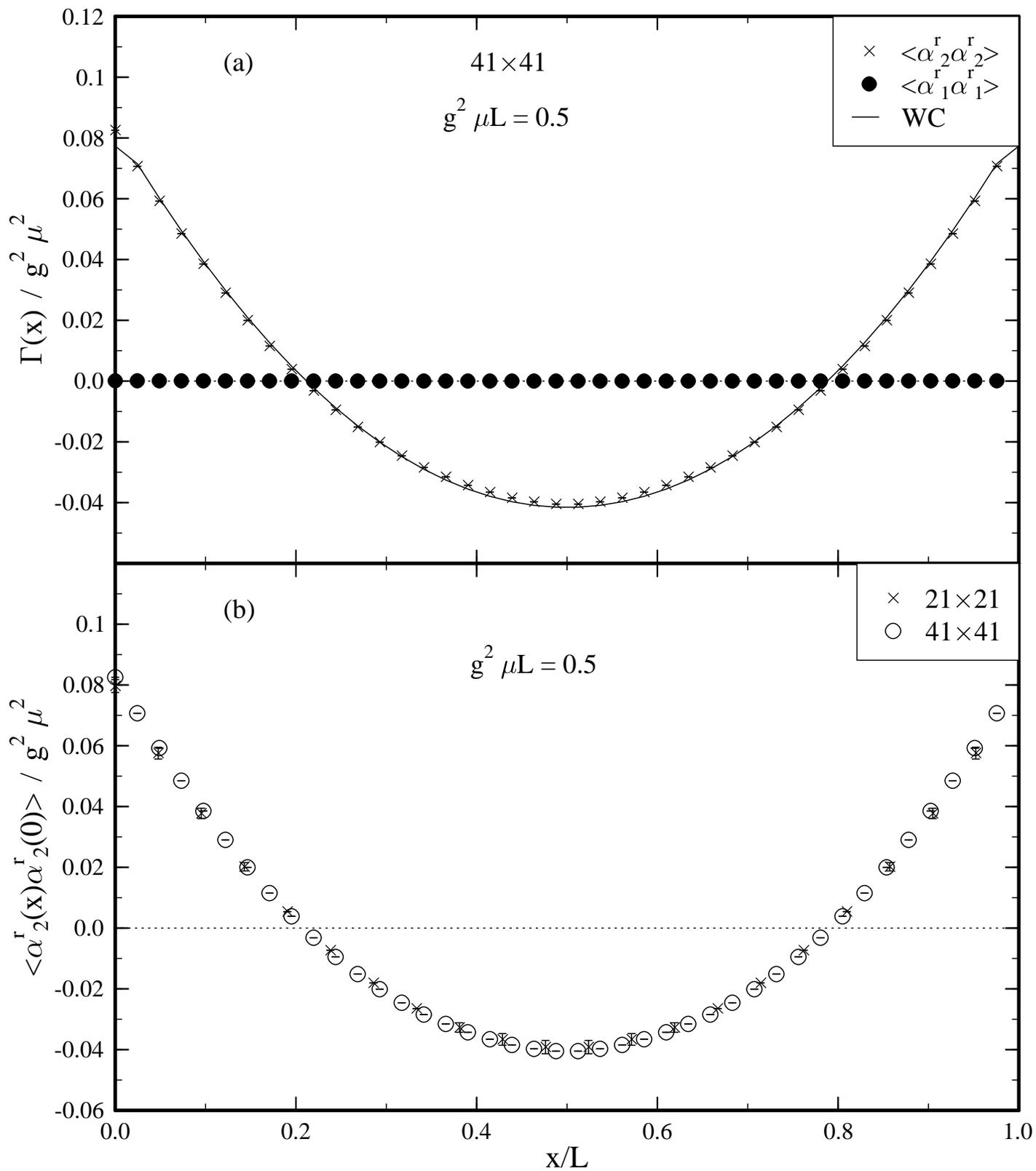
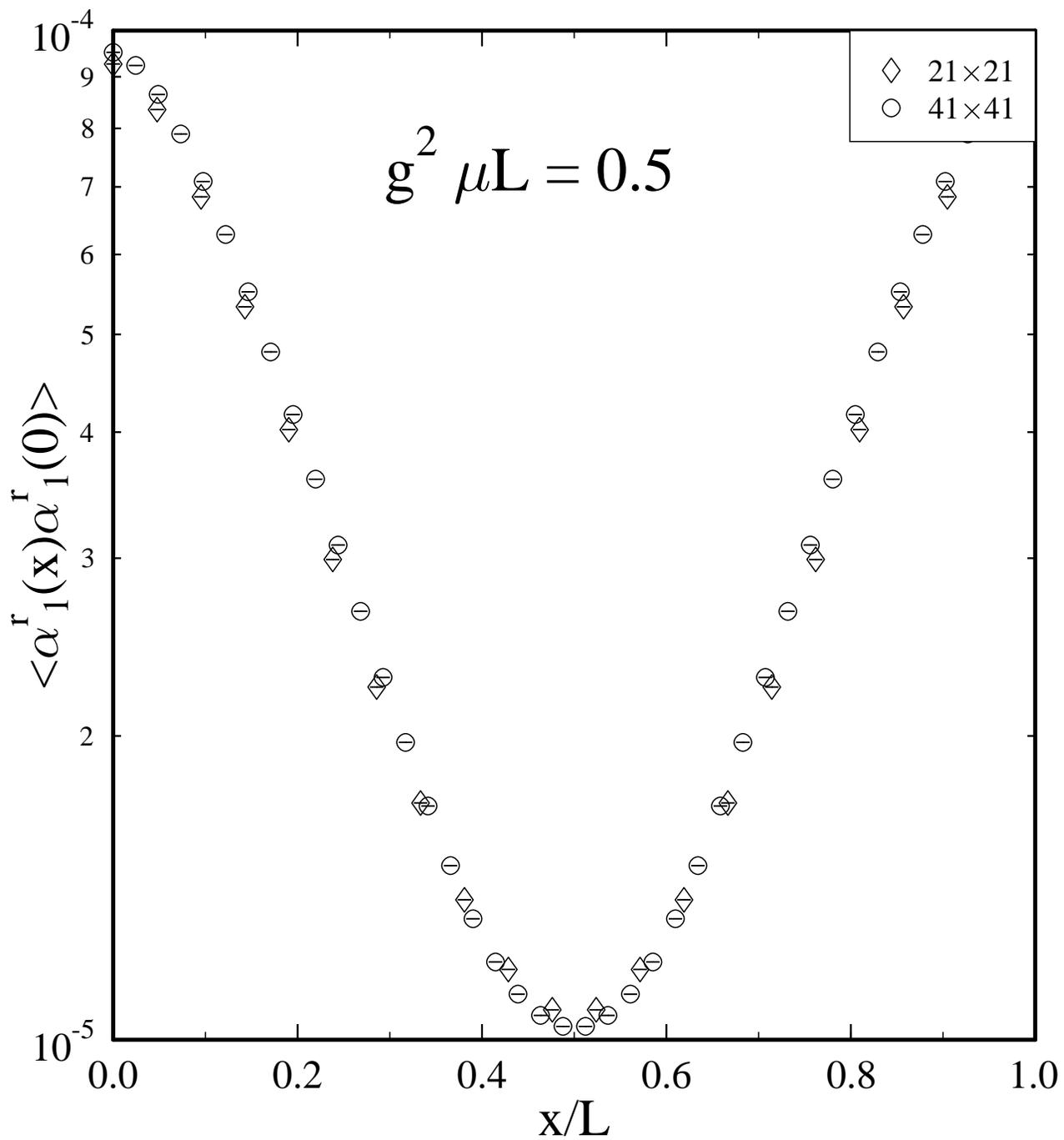
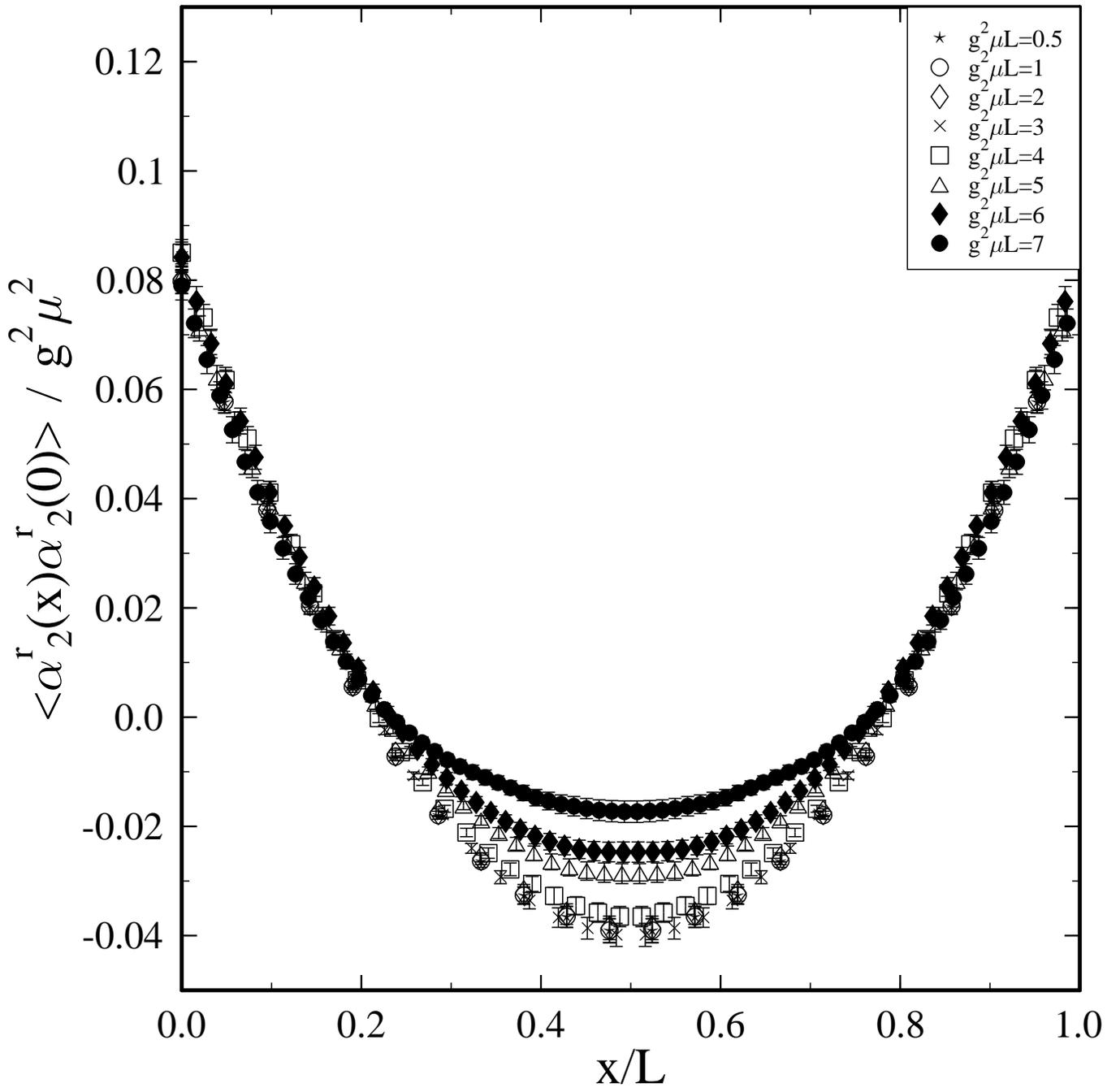
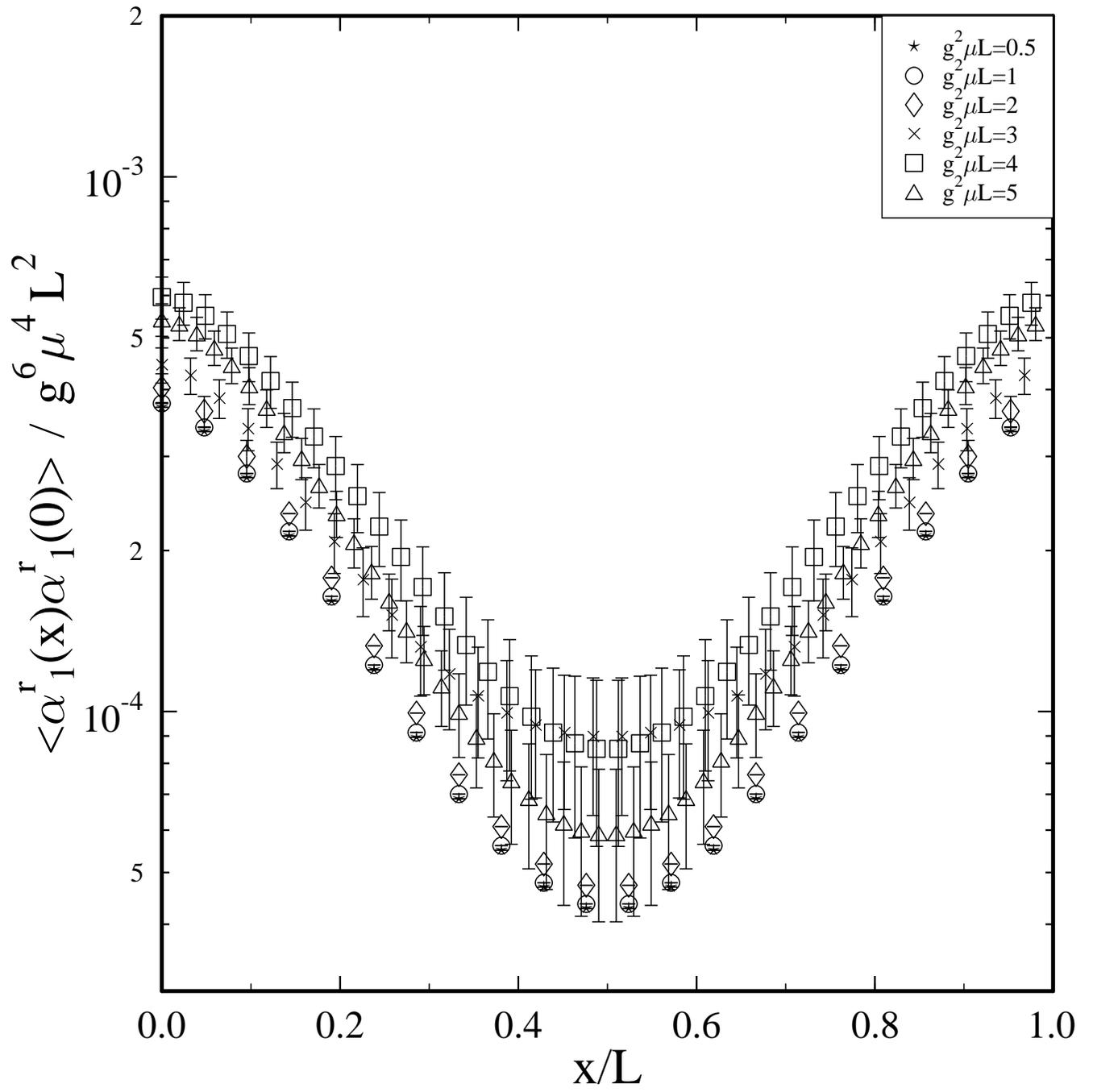
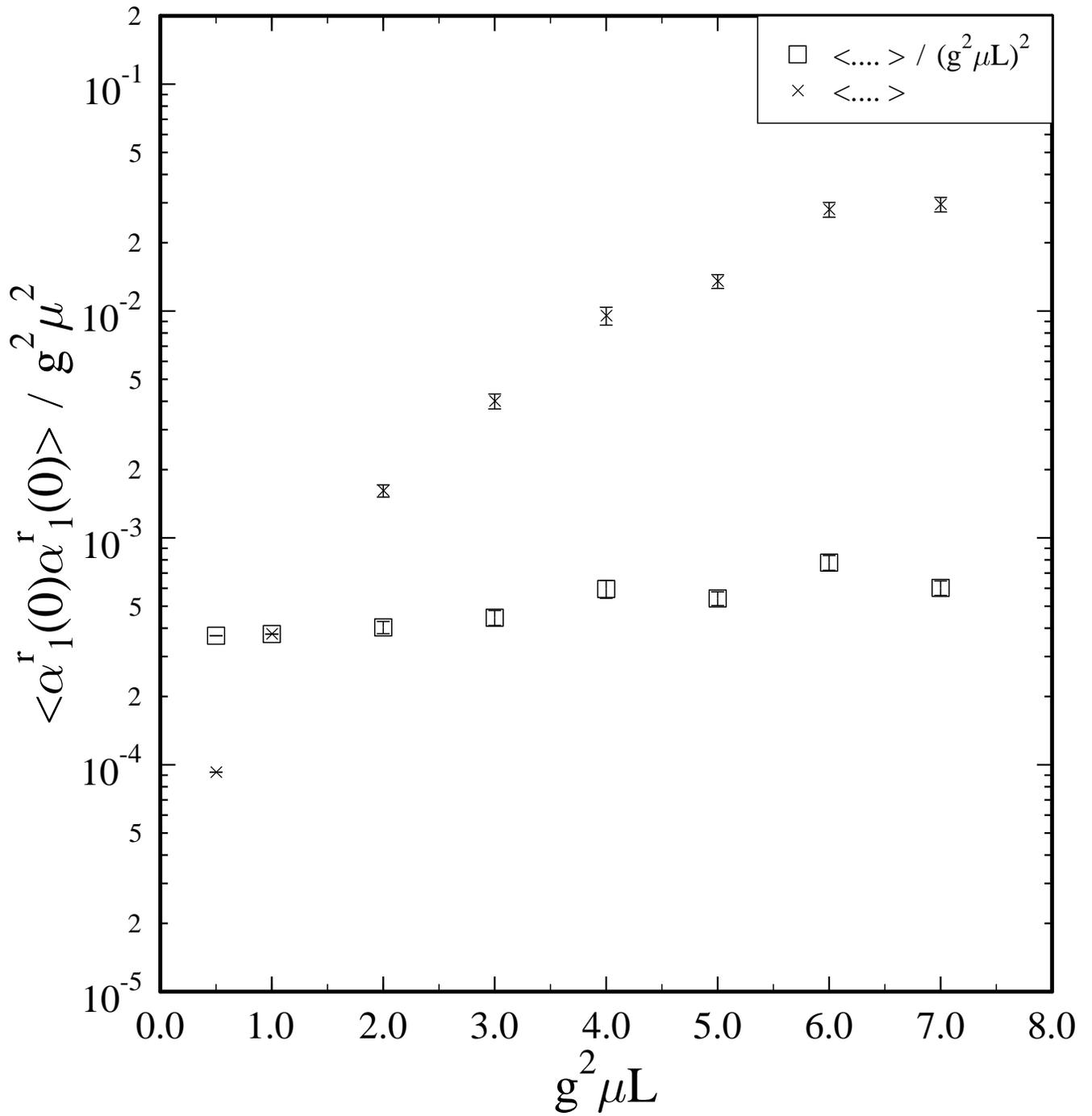
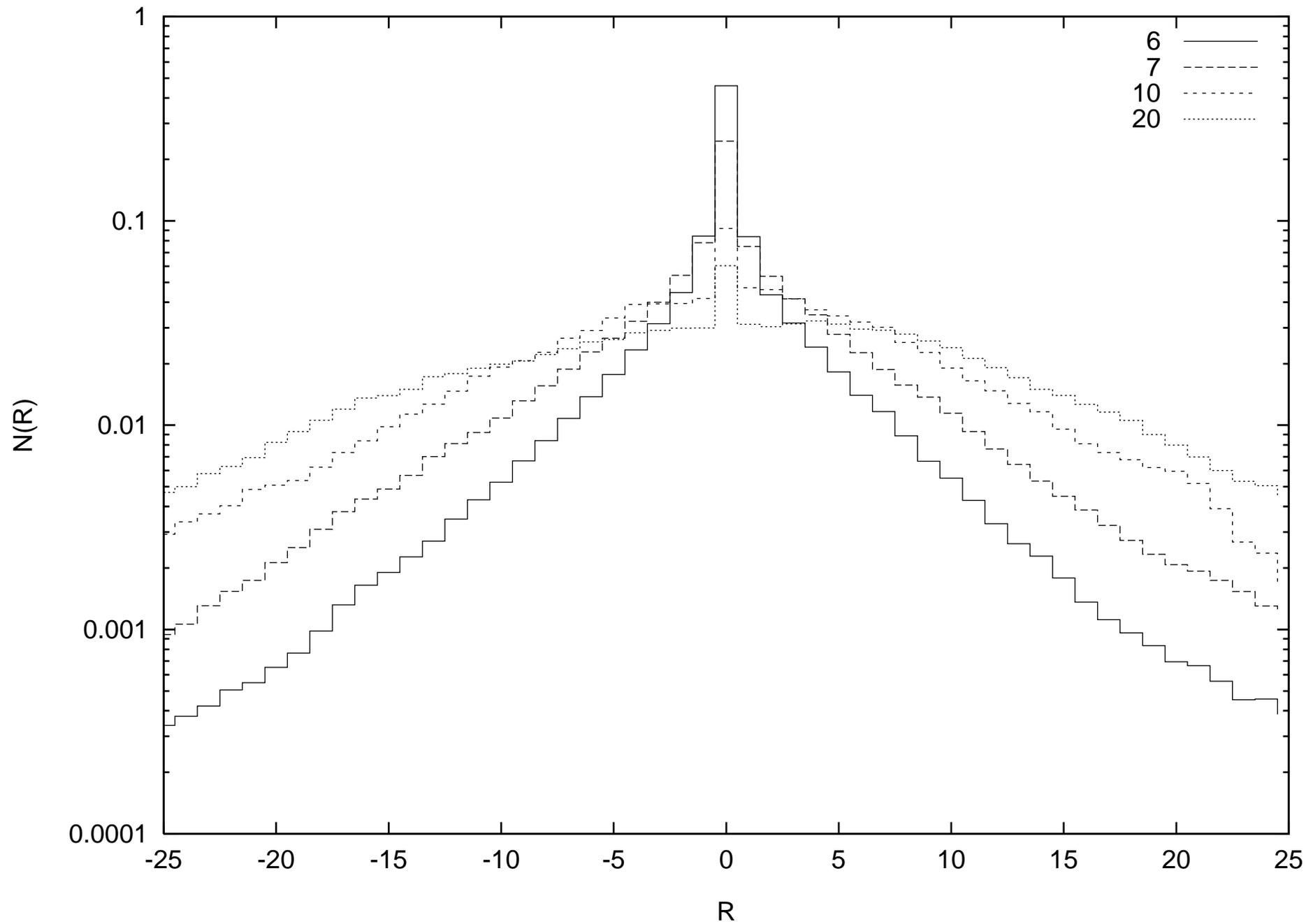
\end{document}